\documentclass[pra,aps,twocolumn,showpacs]{revtex4-2}
\usepackage{graphicx}
\usepackage{epsfig}
\usepackage{color}
\usepackage{subfig}
\begin{document}
\title{Optical chiral properties in a large resonant hybrid photonic cluster.}
\author{A. D'Andrea}
\affiliation{Istituto di Struttura della Materia, CNR-ISM, C.P. 10, 
Monterotondo Stazione, Roma I-00015. Italy. On retirement.}
\date{\today }
\begin{abstract}
Optical chiral properties of a resonant hybrid photonic crystal (RHPC) are computed taking into account spin-orbit effect due to light-hole excitons perfectly confined in 2D quantum wells. The trends of the optical activity, expressed as a ratio between the absorption intensities of the z and xy light-hole polaritons, are obtained by computing the optical response in a rather large N-cluster of elementary cells (N= 34) and for exciton energy, in resonance with a stationery inflection point (SIP). High values of spin-orbit interaction (0.7 eV\AA) produce strong distortions of the optical activity polar curves that, differently, becomes rather isotropic if the experimental value (0.14 eV\AA) is used.
\end{abstract}
\maketitle

\section{INTRODUCTION}
It is well known that meta-materials can shows optical activity due to a special design of building blocks, while semiconductor quantum confined nanostructures (wells, wires and dots) are optically active as grown \cite{1}.
 Since the  functionality of  an optical device is due to the optimization  of bulk and surface optical properties the tailoring of the sample, in order to study a particular optical property,  proceed by optimizing the symmetry of the bulk elementary cell of the resonant hybrid photonic crystal 
%
%
\begin{figure}[t]
\includegraphics[scale=0.35]{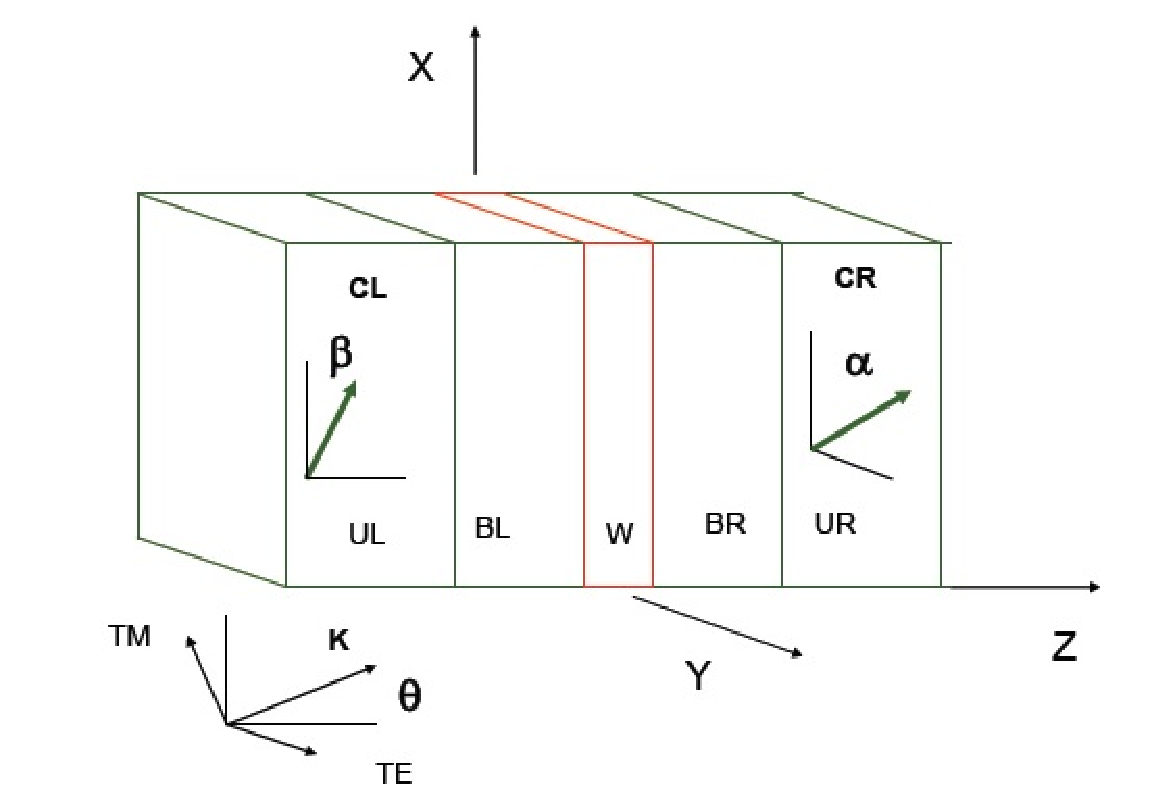}
\caption{The elementary cell of the "minimum model" of 1D Resonant Hybrid Photonic Crystals (RHPC).}
\end{figure}
(RHPC) in the first part of the present work, while   the $N$-cluster optical properties will be optimized in the second part.
In the present work it is considered as meta-material an one-dimensional hybrid (isotropic-anisotropic) photonic crystals (HPC) and, as semiconductor nanostructures, a two-dimensional quantum wells embodied in the isotropic part of the photonic elementary cell \cite{2,3}. 
In recent papers  \cite{2,3}, the Author has studied the conditions under which the so called "minimum model",  of a resonant hybrid photonic crystals (RHPC), can show interactions between light-hole exciton spin orbit  and photon spin orbit effects  for spin-orbit interaction value $\beta _{ex}  = 0.7eV$\AA \cite{3} (five times  greater than the theoretical value for III-V semiconductors: $\beta _{ex}  = 0.14eV$\AA \cite{4}). 
In particular the presence of axial spectral asymmetry namely: $\hbar \omega (k_\parallel  ,K_Z ) \ne \hbar \omega (k_\parallel  , - K_Z )$  \cite{5} in the elementary cell, of an 1D hybrid photonic crystal, could give the presence of a stationary inflection point (SIP) in the dispersion curves \cite{5,6,7} that are at basis of the presence of non-Bloch solution in Maxwell equation and of  the strong deformation  \cite{3,5} of polar symmetry, also under non-resonant condition,   in such kind of hybrid photonic crystals (HPC).  Notice that the condition for the axial spectral asymmetry  imposes certain restrictions on the geometry of the hybrid cell since it is due to the different optical paths that the forward and backward extraordinary rays can cover inside the anisotropic cell \cite{6,7} (the  forward and backward extraordinary rays must have different $\left| {k_{xz} } \right|$ values). On the other hand, the axial spectral asymmetry essentially requires oblique wave propagations  and nonzero $xz$-component of the dielectric permittivity tensor.
Moreover, the former conditions hamper to observe these points at normal incidence in a non-magnetic hybrid photonic crystal since the time reversal symmetry  is conserved.
Obviously, when the transition energy of Wannier exciton is in resonance with the energy of the SIP, in the dispersion curves of HPC, also unidirectional exciton-polariton propagation is present in such kind of  resonant hybrid photonic crystal (RHPC).

%
%
\begin{figure*}
\centering
\subfloat[]{\includegraphics[scale=0.42]{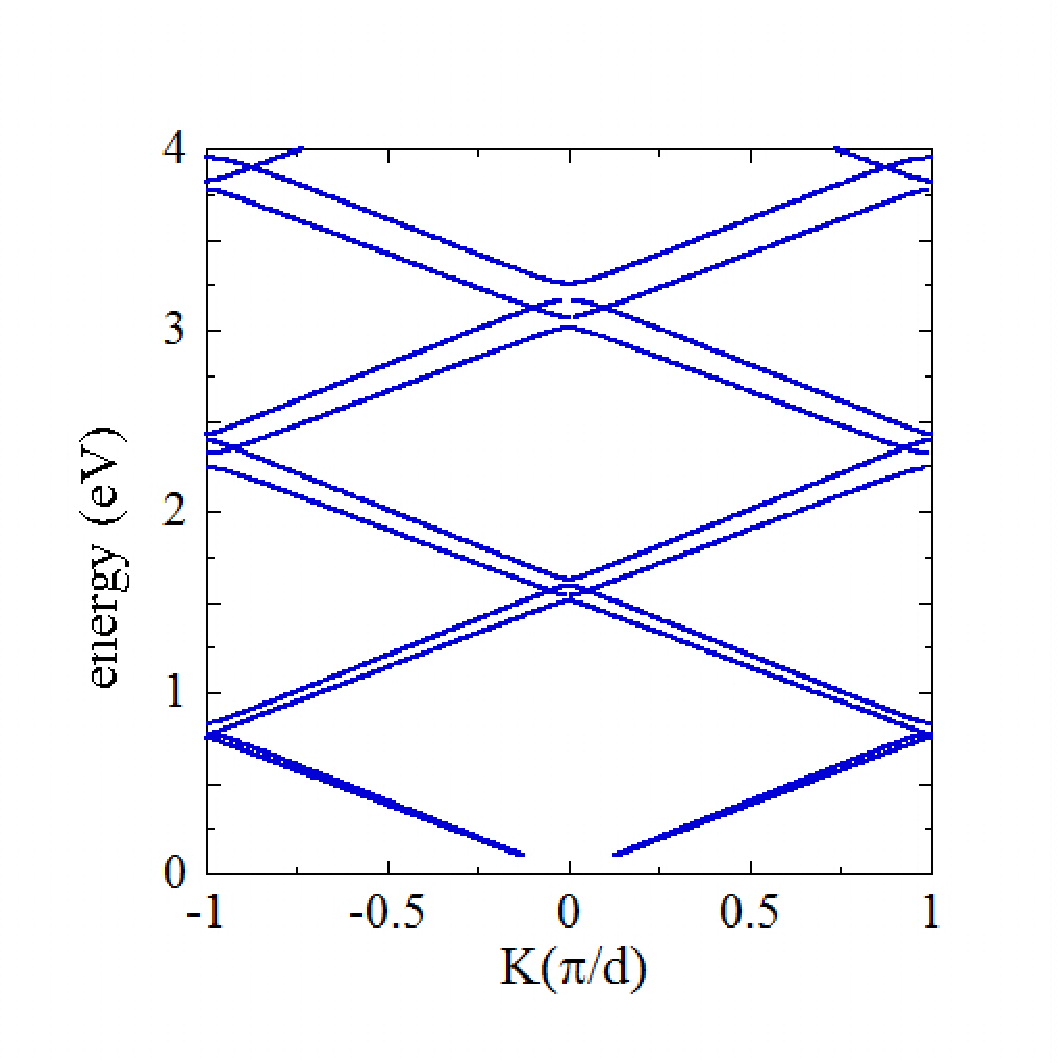}} \qquad          
\subfloat[The same of fig. 2a in enlarged scale]{\includegraphics[scale=0.42]{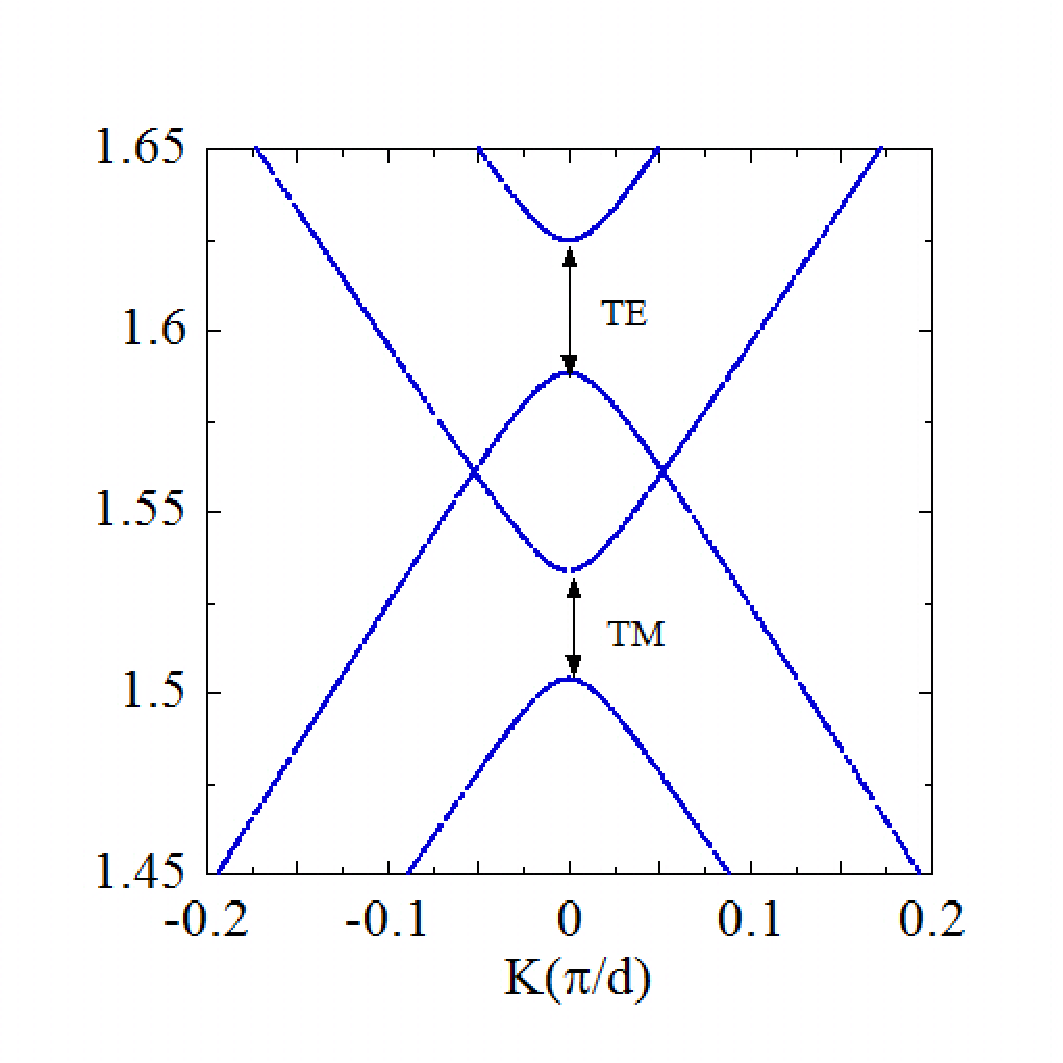}}
\caption{Photonic dispersion curves for  hybrid photonic crystal (HPC) with symmetric elementary cell. $\vartheta  = 20^\circ ;\alpha  = \beta  = 0^\circ $.} 
\end{figure*}
%
%
\begin{figure*}
\centering
\subfloat[]{\includegraphics[scale=0.42]{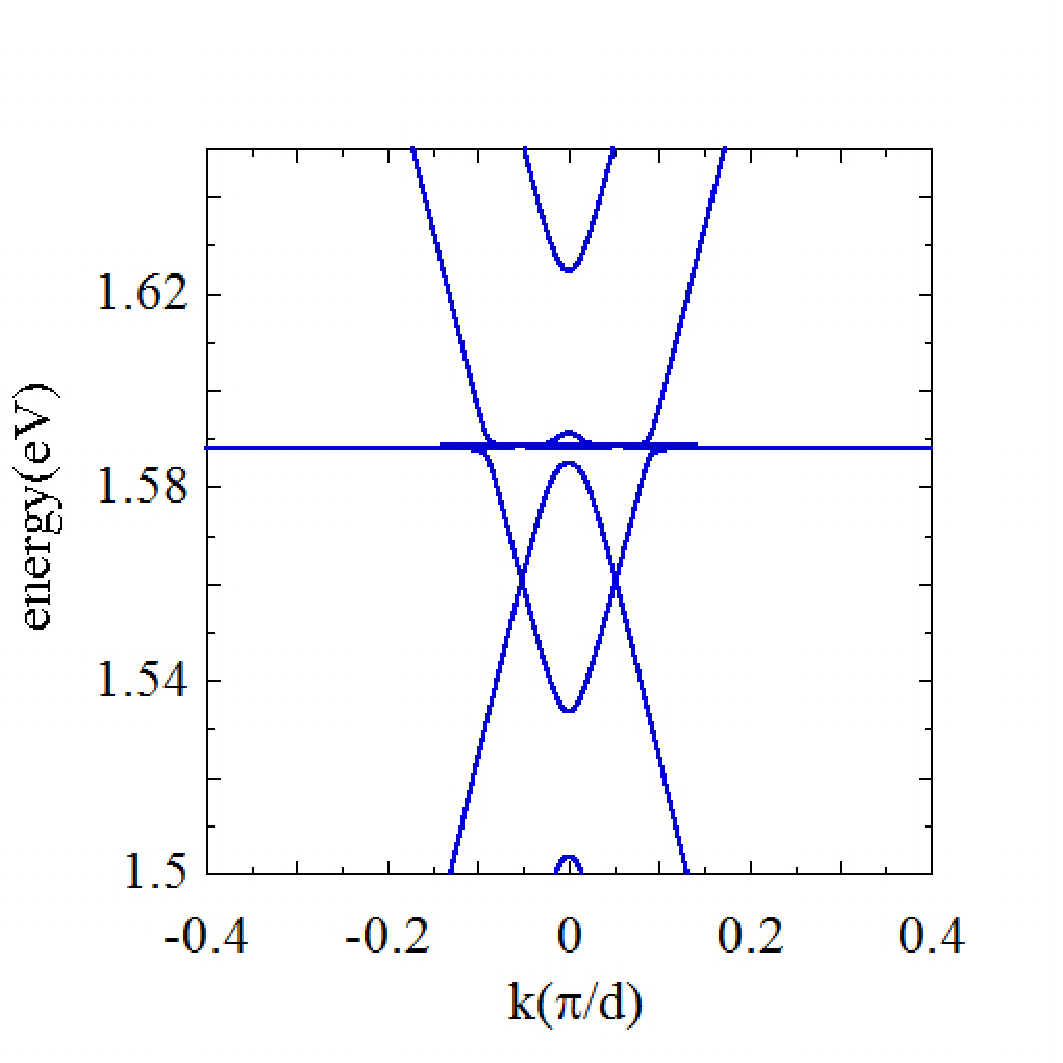}} \qquad          
\subfloat[The same of Fig. 3a  in enlarged  scale close to  $\Gamma $ point.]{\includegraphics[scale=0.42]{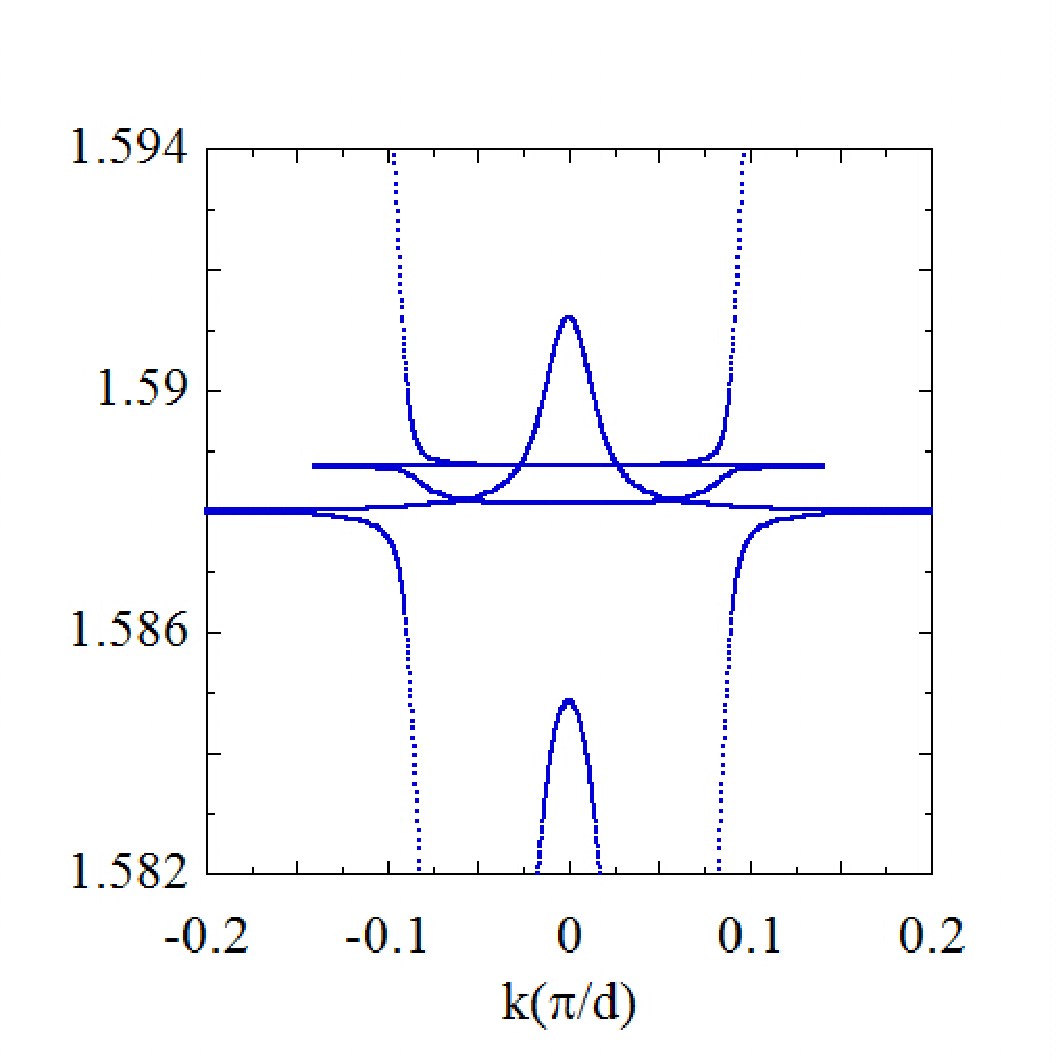}}
\caption{ Polariton dispersion curves for light-hole exciton  energy: $E_{ex} (k_\parallel   = 0) = 1.588$ eV; $\vartheta  = 20^o ,\,\,\alpha  = \beta  = 0^o $.
} 
\end{figure*}
%
%
\begin{figure}[t]
\includegraphics[scale=0.42]{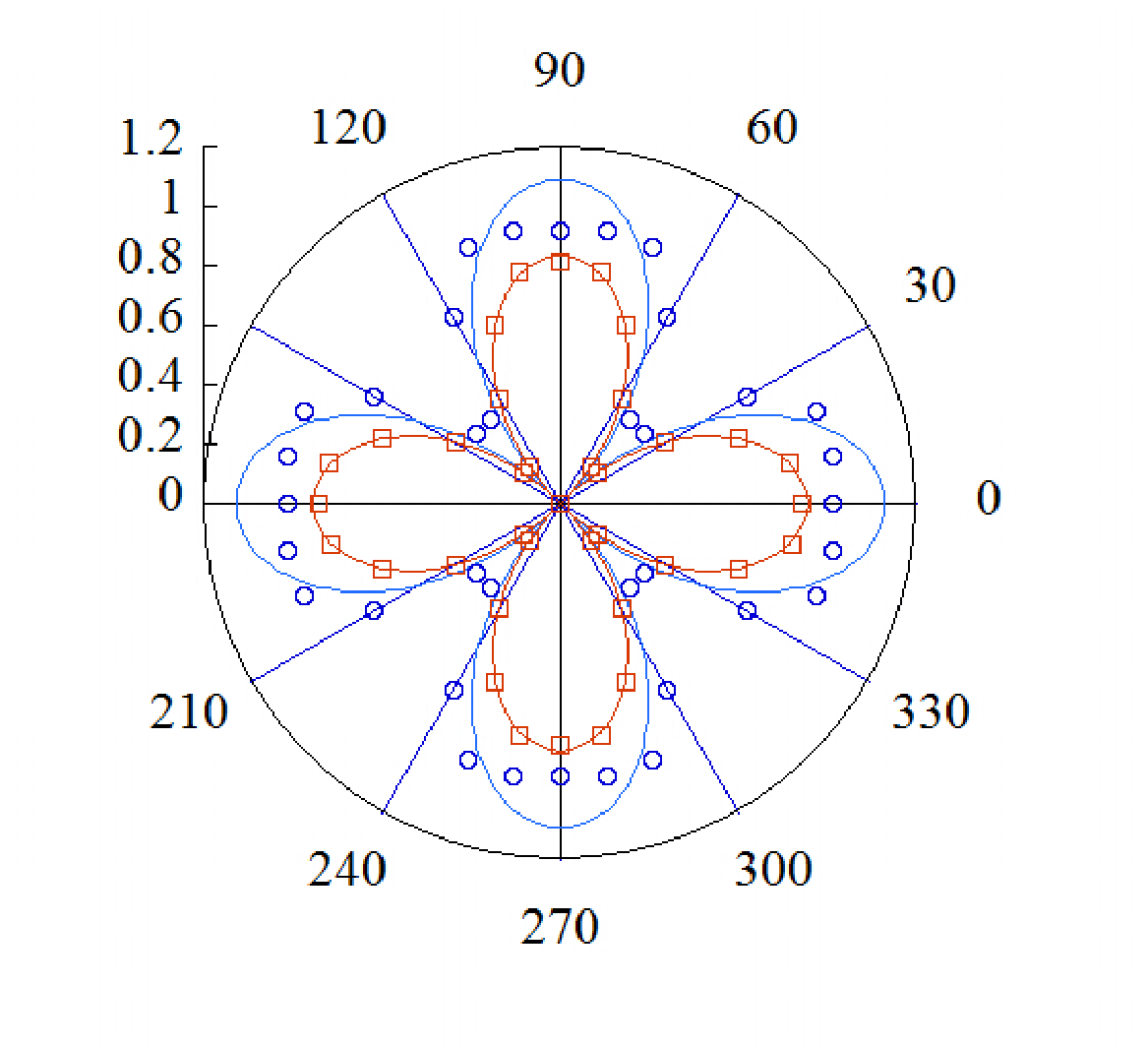}
\caption{Value of  $ I_z /I_\parallel$ computed, in resonance with light-hole exciton energy, for a symmetric N-cluster (N = 34) as a function of $\varphi$ rotation angle. $\beta _{ex}  = 0.7$ eV{\AA} (blue circles) and $\beta _{ex}  = 0.14$ eV{\AA}  (red squares). $\theta  = 20^o ,\;\alpha  = \beta  = 0^o $. Solid lines are a fit by  $\left| {A\cos 2\varphi } \right|$.}
\end{figure}
In order to enhance the tensorial properties of the anisotropic layers the background dielectric contrast, between isotropic and anisotropic layers, has to be chosen as small as possible, therefore for anisotropic and isotropic  layers are chosen: $\varepsilon _\parallel   - \varepsilon _ \bot   \approx 1$    and: $\varepsilon _b  \approx \left( {\varepsilon _\parallel   + \varepsilon _ \bot  } \right)/2$  respectively \cite{3} because these conditions avoid the presence of an independent from the polarization photon energy gap,  at the lowest photon energies.  Notice that while in the simple two-layer photonic crystals the absolute photonic gap is  usually achieved by modulating the background dielectric contrast, in hybrid photonic crystals the optical gap, independent from polarization, is obtained by more sophisticated tensorial properties as they it will be shown in the following.
The strategy adopted for the tailoring of RHPC is based on starting from an elementary cell  with fully symmetric dispersion curves, at non-normal incidence, where the two optical $C$-axes of the anisotropic layers are both parallel to $x$-axis (see Fig. 1) and by  increasing its complexity  step by step by $C$-axes rotation.
As it is well known,  the strained materials of  2D quantum well allows to compute separately  heavy and light-hole optical response and moreover,  since each layer of the HPC shows a single dielectric tensor  (except the quantum well),  this allows you to impose the normal Maxwell boundary conditions at each interfaces of the heterostructure,  while chiral property is shown at the boundary of the total elementary cell: 
$\vec D(\omega ) = \mathord{\buildrel{\lower3pt\hbox{$\scriptscriptstyle\leftrightarrow$}} 
\over \varepsilon } \,\vec E(\omega ) + \,\mathord{\buildrel{\lower3pt\hbox{$\scriptscriptstyle\leftrightarrow$}} 
\over \alpha } \,\vec \nabla  \times \vec E(\omega )$
\section{The MINIMUM MODEL}
The studied multilayers have spatial periodicity $d = 400nm$  along the Cartesian $z$ axis and their photonic elementary cell is composed by isotropic (cubic $L_C=d/2 $)  and anisotropic (uniaxial $L_{UL}+L_{UR}=d/2$) layers.  In turn, the anisotropic part is composed of two uniaxial layers, placed at the two ends of the cell,  both of  $d/4$  thickness but  with different optical $C$-axes orientations  (see Fig. 1). Once fixed the reference frame such that the wave vector of incident wave   lies on the $(x,z)$ plane, the  $C$-axis of the left uniaxial layer lies on this plane of incidence  while the $C$-axis of the right uniaxial layer  lies on the $(x,y)$  reflecting plane.  Finally, the isotropic layer consists of a perfectly confined 2D quantum well (QW) with well thickness $L_W  = 10nm$ .

\section{RADIATION-MATTER INTERACTION: DISPERSION CURVES}
The components of  dielectric tensor in-plane $(x,y)$  and along $z$-axis are respectively: $\varepsilon _\parallel   = 4.9$  and $\varepsilon _ \bot   = 3.9$  while the background dielectric constant of quantum well and barriers are $\varepsilon _w  = 8.4$  and $\varepsilon _b  = 4.4$.respectively. Notice that the above values of the optical parameter, adopted in the calculation, are  rather close to the hybrid system $YVO_4 /GaAs$.
The  other  parameter values of the exciton-polariton model are:  the Kane's energy of light-exciton transition is $E_K  = 23\,eV$, while $a_{B}  = 8.047nm$  is the exciton Bohr radius, $M = 0.59\,m_e $  is the exciton center-of-mass and the non-radiative exciton broadening is $\Gamma _{NR}  = 0.1meV$.
%
%
\begin{widetext}
\begin{equation}
\left\{ \begin{array}{l}
 \left( {\omega _{ex}  - \omega } \right)P_x (z,\omega ) - i\frac{{\beta _{ex} }}{\hbar }\,\left( {\frac{{S_\parallel ^o }}{{S_ \bot ^o }}} \right)^{1/2} q_y \,P_z (z,\omega ) = S_\parallel ^o \Psi ^ *  (z)\int\limits_{ - L_w /2}^{L_w /2} {dz'} \,\Psi (z')\,E_x (z',\omega ) \\ 
 \left( {\omega _{ex}  - \omega } \right)P_y (z,\omega ) - i\frac{{\beta _{ex} }}{\hbar }\,\left( {\frac{{S_\parallel ^o }}{{S_ \bot ^o }}} \right)^{1/2} q_x \,P_z (z,\omega ) = S_\parallel ^o \Psi ^ *  (z)\int\limits_{ - L_w /2}^{L_w /2} {dZ'} \,\Psi (z')\,E_y (z',\omega ) \\ 
 \left( {\omega _{ex}  - \omega } \right)P_z (z,\omega ) + i\frac{{\beta _{ex} }}{\hbar }\,\left( {\frac{{S_ \bot ^o }}{{S_\parallel ^o }}} \right)^{1/2} \left[ {q_y \,P_x (z,\omega ) + q_x \,P_y (z,\omega )} \right]\, = S_ \bot ^o \Psi ^ *  (z)\int\limits_{ - L_w /2}^{L_w /2} {dZ'} \,\Psi (z')\,E_z (z',\omega ) \\ 
 \end{array} \right.
\end{equation}
\end{widetext}
where $\Psi (z)$ is the exciton wave function.  The two coefficients $(S^0 )$ with dipole oriented along the $z$-axis $(S_ \bot ^o )$ and  on quantum well plane $(S_\parallel ^o )$ respectively,  are in the ratio $S_ \bot ^o /S_\parallel ^o  = 4$  as already described in Ref.  \cite{3}.  
In the polariton framework the dipole transition energy is given by: 
$\hbar (\omega _\parallel   - \omega ) = \hbar (\omega _ \bot   - \omega ) = \hbar (\omega _{ex}  - \omega )$
since  the difference between in-plane and orthogonal energies $\hbar (\omega _\parallel   - \omega ) \ne \hbar (\omega _ \bot   - \omega )$  comes from the real part of the radiative self-energy \cite{2,3}. Therefore, by dividing for the energy denominator: $\hbar \left( {\omega _{ex}  - \omega } \right) = E_{ex} (q_\parallel   = 0) - \hbar \omega  - i\,\Gamma _{NR} $
 we obtain:
%
%
\begin{widetext}
\begin{equation}
\left\{ \begin{array}{l}
 P_x (z,\omega ) = i\frac{{\beta _{ex} }}{{\hbar (\omega _{ex}  - \omega )}}\,\left( {\frac{{S_\parallel ^o }}{{S_ \bot ^o }}} \right)^{1/2} q_y \,P_z (z,\omega ) + \frac{{S_\parallel ^o }}{{\hbar (\omega _{ex}  - \omega )}}\Psi ^ *  (z)\int\limits_{ - L_w /2}^{L_w /2} {dZ'} \,\Psi (z')\,E_x (z',\omega ) \\ 
 P_y (z,\omega ) = i\frac{{\beta _{ex} }}{{\hbar (\omega _{ex}  - \omega )}}\,\left( {\frac{{S_\parallel ^o }}{{S_ \bot ^o }}} \right)^{1/2} q_x \,P_z (z,\omega ) + \frac{{S_\parallel ^o }}{{\hbar (\omega _{ex}  - \omega )}}\Psi ^ *  (z)\int\limits_{ - L_w /2}^{L_w /2} {dZ'} \,\Psi (z')\,E_y (Z',\omega ) \\ 
 P_z (z,\omega ) =  - i\frac{{\beta _{ex} }}{{\hbar (\omega _{ex}  - \omega )}}\,\left( {\frac{{S_ \bot ^o }}{{S_\parallel ^o }}} \right)^{1/2} \left[ {q_y \,P_x (z,\omega ) + q_x \,P_y (z,\omega )} \right]\, + \frac{{S_ \bot ^o }}{{\hbar (\omega _{ex}  - \omega )}}\Psi ^ *  (z)\int\limits_{ - L_w /2}^{L_w /2} {dZ'} \,\Psi (z')\,E_z (z',\omega )\quad  \\ 
 \end{array} \right.
\end{equation}
\end{widetext}
Moreover, assuming the incidence plane $(x,z)$ we will have to introduce a new  angle $(\varphi)$ in order to take into account the in-plane rotation of the intrinsic Cartesian axes $\left( {\tilde x,\tilde y,\tilde z} \right)$ orientation: $\tilde x\parallel \left[ {100} \right]$,  $\tilde y\parallel \left[ {010} \right]$ and $\tilde z\parallel \left[ {001} \right]$ of the elementary cell of the semiconductor material with respect to the Cartesian axes of the non-normal  incidence wave geometry and the uniaxial dielectric tensor.
%
%
\begin{figure*}
\centering
\subfloat[$\alpha  = 0^\circ $, $\beta  = 30^\circ $.
  The same  for $\beta  = 0^\circ $  are also shown for sake of comparison.]{\includegraphics[scale=0.42]{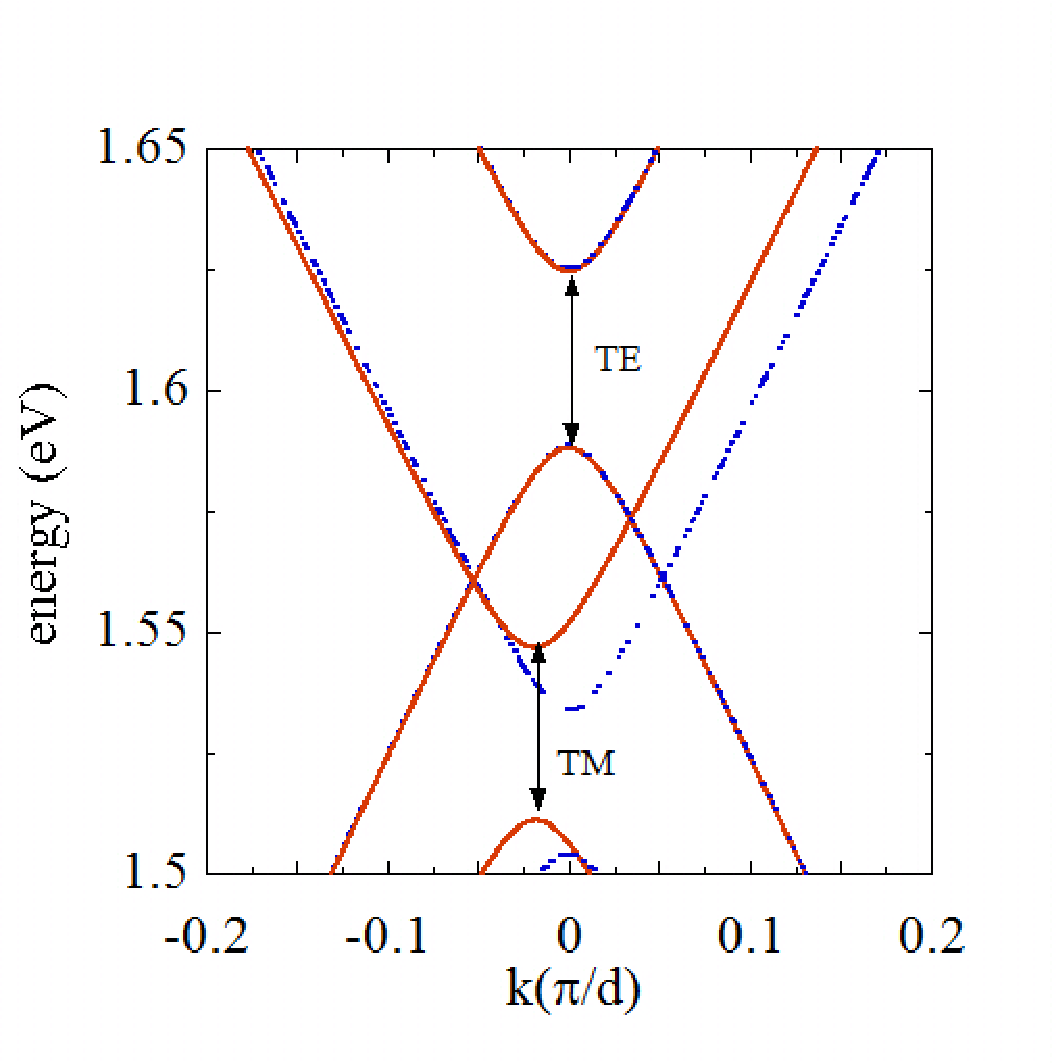}} \qquad          
\subfloat[$\alpha  = 45^o $, $\beta  = 30^o $. The same for $\alpha  = 0^o $ is also shown for sake of comparison.]{\includegraphics[scale=0.42]{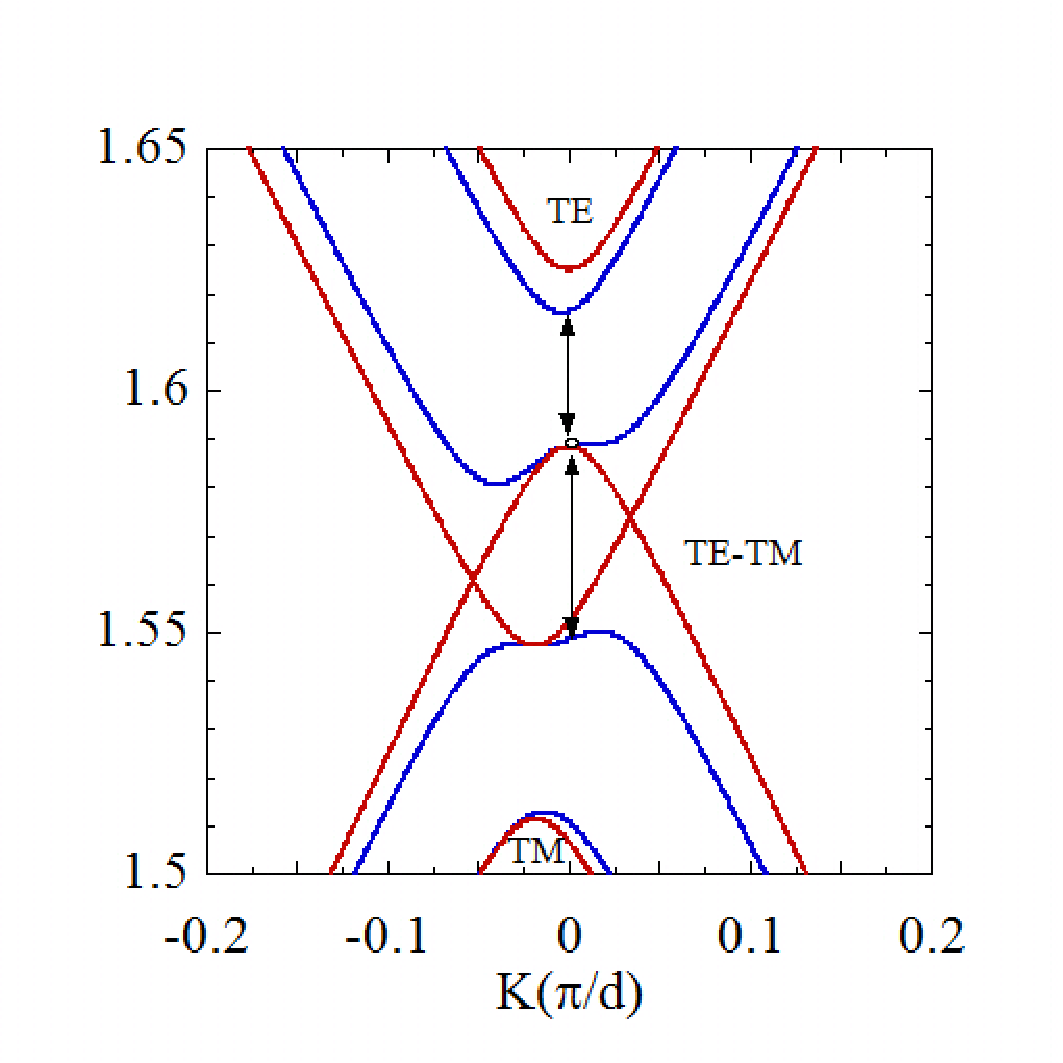}}
\caption{Photonic dispersion curves computed  for  asymmetric elementary cells obtained by  rotation of the optical  $C$-axes. $\beta _{ex}  = 0.14$ eV{\AA}.} 
\end{figure*}
%
%
\begin{figure*}
\centering
\subfloat[$\alpha  = 0^\circ $, $\beta  = 30^\circ$.]{\includegraphics[scale=0.42]{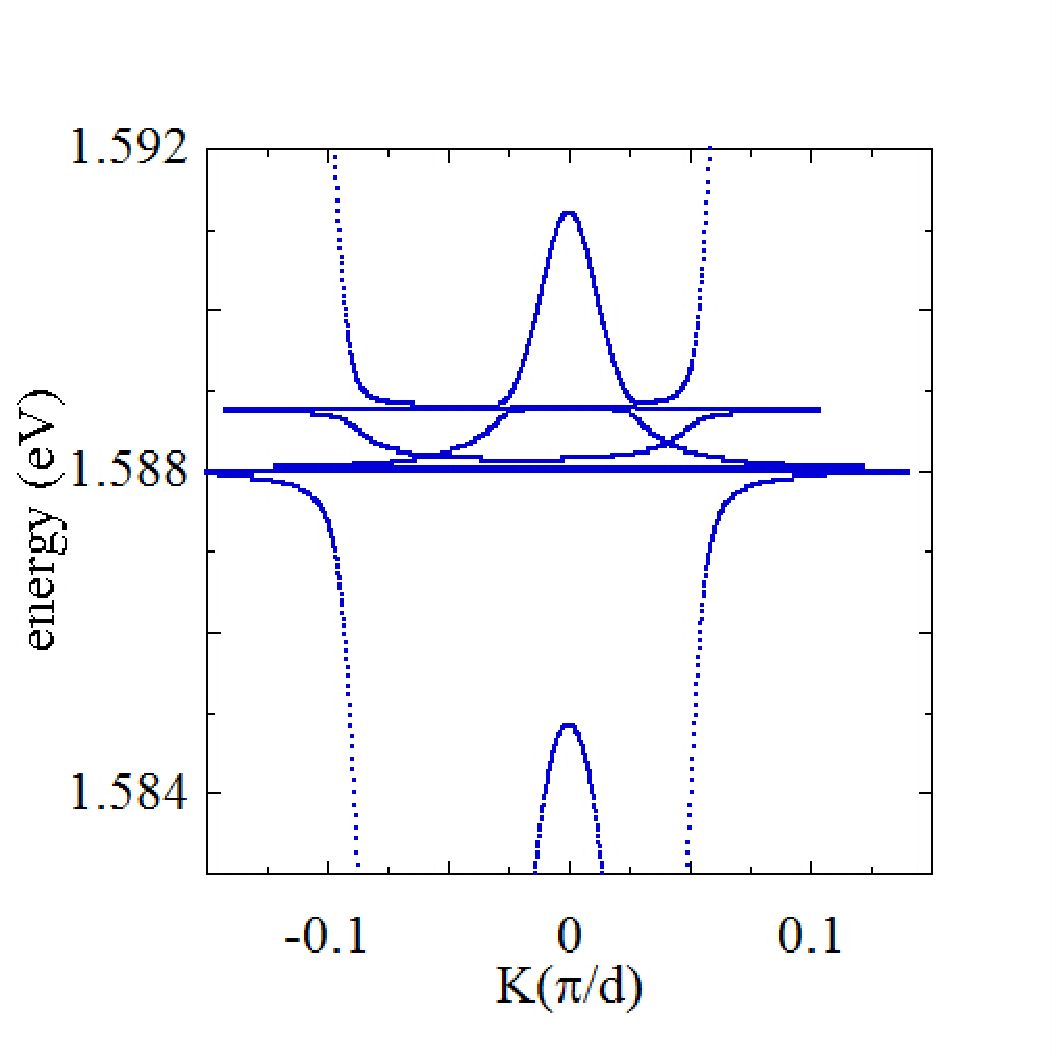}} \qquad          
\subfloat[$\alpha  = 45^\circ $, $\beta  = 30^circ $.]{\includegraphics[scale=0.42]{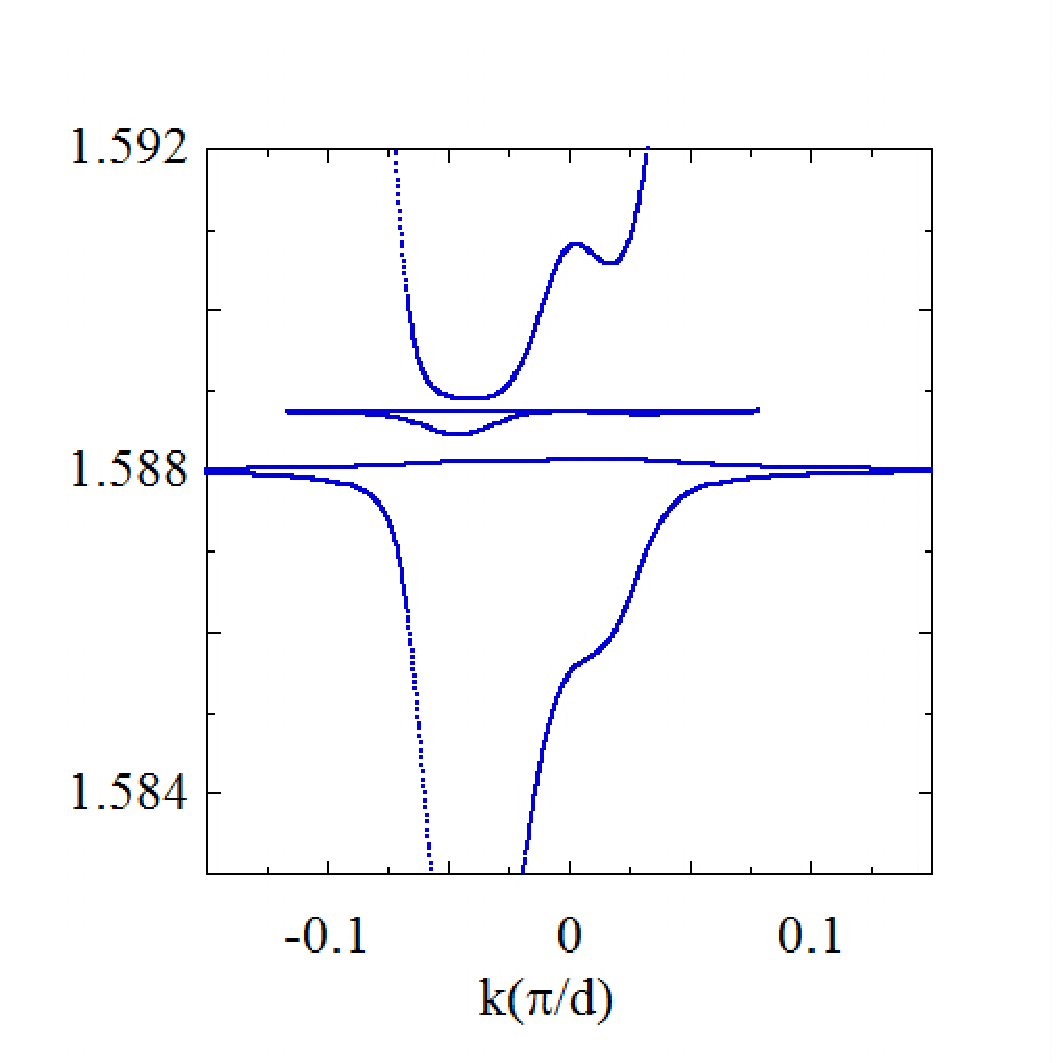}}
\caption{Polaritonic dispersion curves computed  for an asymmetric cell obtained by  rotation of the optical  $C$-axes. $\beta _{ex}  = 0.14$ eV{\AA}.} 
\end{figure*}

Non-normal  $(\vartheta  = 20^\circ )$ photon dispersion curves for TE and TM polarizations are shown in Fig. 2(a).  Notice that, in this case,  no curves appear corresponding to the mixing of the two above polarizations.  Also the photonic gaps, for the two polarizations,  result well separated in energy, as  it  can be seen  in Fig. 2(b) where the curves are shown on enlarged scale,    close to the $\Gamma $ point, and where two symmetric cross points (like Dirac points $ \approx 1.58eV$) appear in the picture.
%
%
\begin{figure*}
\centering
\subfloat[Computed in resonance with SIP energy (red squares) and out-of-resonance (green triangles). The symmetric case is also shown for sake of comparison (blue circles). Spin-orbit interaction $\beta _{ex}  = 0.7$ eV{\AA}
 ]{\includegraphics[scale=0.42]{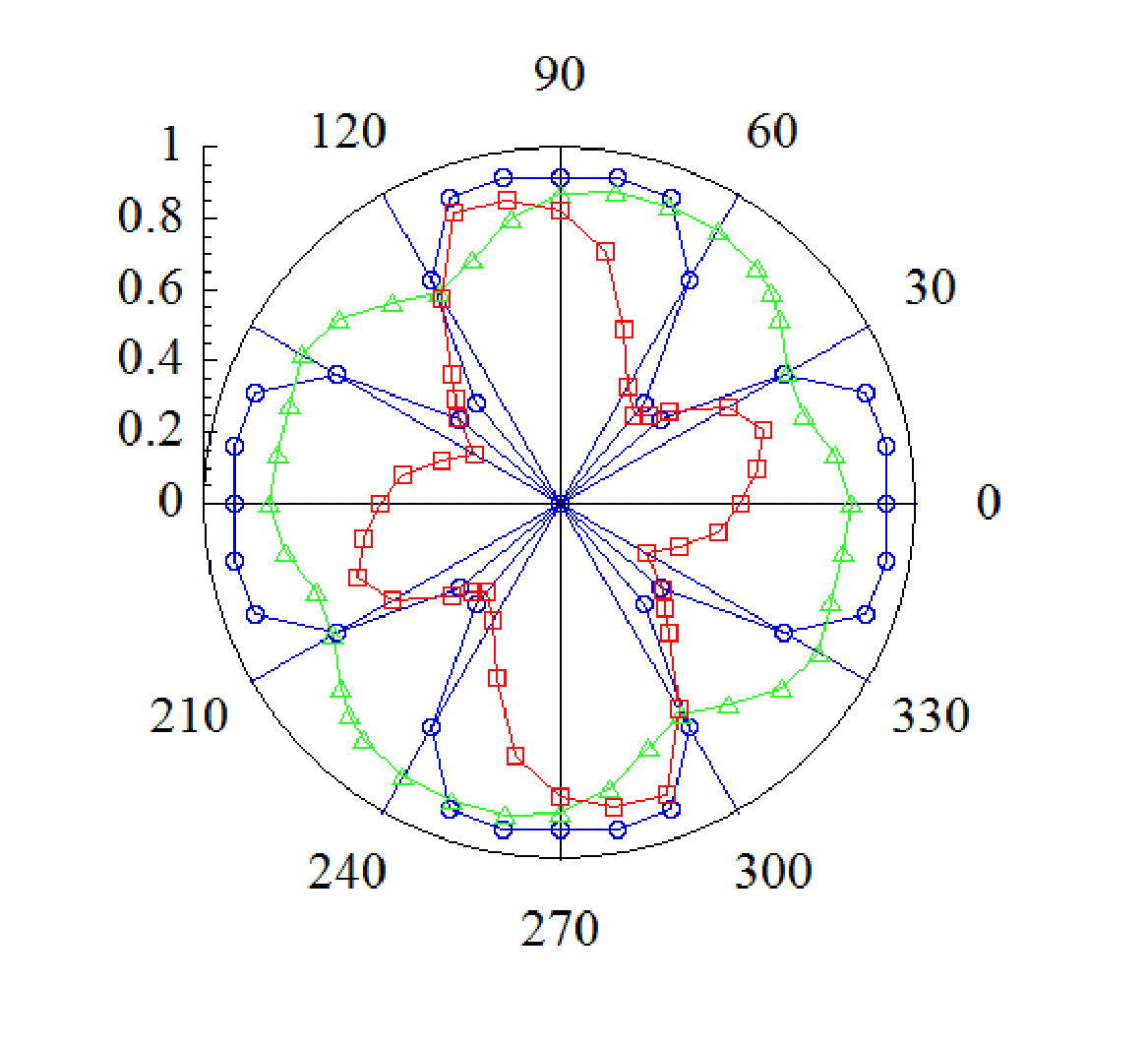}} \qquad          
\subfloat[ Computed in resonance with SIP energy (red squares). The symmetric case is also shown for sake of comparison (blue circles).  Spin-orbit interaction $\beta _{ex}  = 0.14$ eV{\AA}. ]{\includegraphics[scale=0.42]{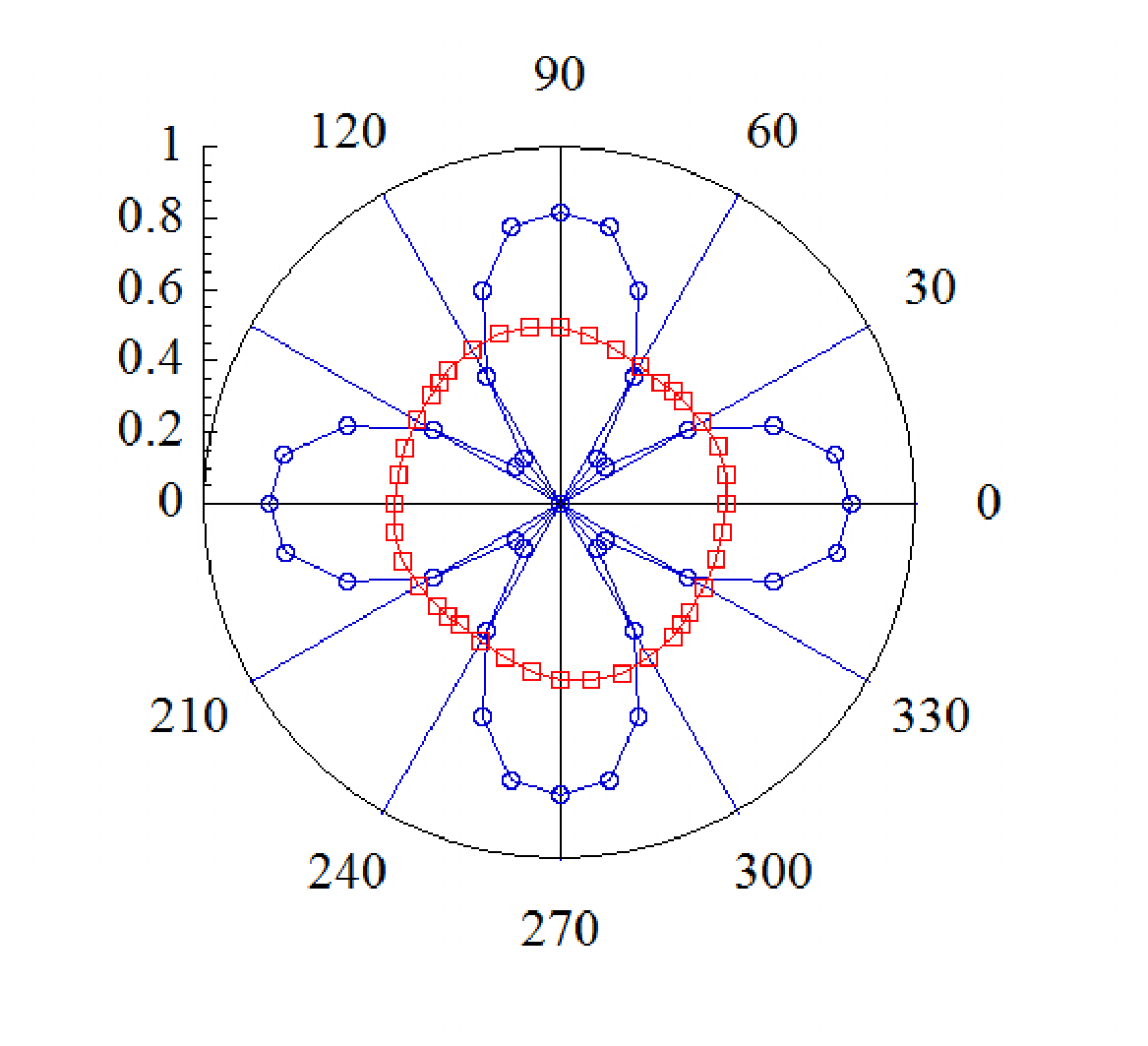}}
\caption{Value of  $I_z /I_\parallel$ computed for an asymmetric $N$-cluster as a function of $\varphi$ rotation angle. Solid lines are just an aid to eyes.} 
\end{figure*}
Finally, the fourth order in exciton-spin-orbit interaction $(\beta _{ex} )$  Maxwell equations of  the electric field in  Cartesian components of radiation-matter interaction are solved  numerically in the polariton framework, by using Green function method  and effective mass approximation  \cite{9}.
The here used non-local polarization tensor ($P$), of radiation-matter interaction is a polariton generalization of that proposed by L.V. Kotova et al. \cite{1} and it is fully discussed in in Refs. \cite{2,3}.

Now, let me consider the dispersion curves in the same energy range of  Fig. 2(b) where also a light-hole exciton, with dipole transition energy: $E_{ex} (k_\parallel   = 0) = 1.588\,eV$, in coincidence with the bottom  of the TE photonic gap, is present in the system. Notice that under the former condition,  for  non-normal incidence TE-polarized wave no TM contribution ($z$-polaritons) will be appears in the optical response if the spin-orbit interaction is neglected $(\beta _{ex}  = 0)$.
The results of the computation are given in Fig. 3(a) and in enlarged scale in Fig. 3(b). Notice that also in this case the symmetry observed in Figs. 2(a) and 2(b) for forward-backward photon propagation is conserved as shown in the two figures. Moreover, two different exciton-polariton energies are present because the  light-hole polariton  posses both in-plane $(1.5879 eV)$ and $z$-polarized $(1.5887 eV)$ components.
Now, let us introduce the spin-orbit effect in the non-local polarization of the light-hole exciton dipole transition as discussed in Refs. \cite{2,3}.  Obviously, spin-orbit effect introduces a non negligible $z$-polariton in the absorbance. Moreover,  we define the Cartesian axes of the semiconductor elementary cell, namely:   $\tilde x \equiv \left[ {100} \right]$ , $\tilde y \equiv \left[ {010} \right]$ and $\tilde z \equiv \left[ {001} \right] = z$, $\varphi $  the angle  between the plane of incidence and  the axis $\left[ {100} \right]$.
In the present calculation as exciton spin-orbit interaction  is used $\beta _{ex}  = 0.14eV$\AA, a value    close to the III-V semiconductors \cite{10} and  also a five times greater value $\beta _{ex}  = 0.7eV$\AA. Notice that while dispersion curves, computed close to the  exciton energy transition and for the lower value of $\beta _{ex} $, remain symmetrical with respect to the forward-backward propagation (see Fig.4(b) in  Ref. \cite{3}) for greater $\beta _{ex} $  value they become asymmetrical (see Fig.4(a) in Ref. \cite{3}) underlining the presence of an intrinsic not negligible optical activity,  as already discussed in the same reference.

\section{OPICAL RESPONSE IN A LARGE N-CLUSTER}
Now let me consider a cluster of $N$-elementary cells of Fig.1, symmetrical with respect the forward-backward wave propagation and for $N$ as large as to show a well formed TE optical gap (see Fig. 2(b)) observed on the optical response under TE oblique  ($\vartheta  = 20^\circ $)   incident wave, this is obtained for $N=34$ elementary cells. 
Therefore, for the optical response computed for TE-polarized non-normal wave of incidence and close to the edge of  TE polarized optical gap, the presence of not negligible small $z$-polariton peak is a fingerprint that there is a TM contribution in the total electric field that, in the present case, it can only be  due to the exciton spin-orbit effect.
In Fig. 4 the ratio between the intensity of $z$ and in-plane polariton absorbance peaks  $(I_z /I_\parallel  )$, computed in resonance with the light-hole exciton energy, is shown for two  values of spin-orbit interaction, namely: $\beta _{ex}  = 0.7eV$\AA    (blue circles) and $\beta _{ex}  = 0.14eV$\AA   (red squares) as function of the angle of the rotation angle $\varphi $  of the elementary cell.  From Fig. 4 clearly appears that, for both the employed $\beta _{ex} $  values, the above ratio is periodic of $\pi /2$  and vanishes if $\varphi $   is an odd multiple of $\pi /4$.  Notice that the present results, obtained by using $\beta _{ex}  = 0.14eV$\AA,  very well reproduce the trend of the experimental circular polarization degree showed in Fig. 3 of Ref. \cite{1}.
The different behavior observed in Fig. 4 for the two spin-orbit interaction values may be explained by considering that, for smaller $\beta _{ex} $   values, the coefficients of both the third and fourth order terms of quartic Maxwell's equation, for the radiation-matter interaction \cite{2,3}, become negligibly small but, differently,  their contribution becomes more relevant when larger values of  $\beta _{ex} $ are used.

In order to obtain an HRPC optically active  the optical $C$-axes of the left and right  uniaxial layers have to be oriented forming an angle  $\alpha $  with the $x$ axis on the plane of incidence $(x,z)$ and an angle $\beta $  with the $x$ axis in-plane $(x,y)$  respectively (see Fig. 1).  Consequently the two corresponding dielectric tensors   will have the following form:
%
%
\begin{equation}
\varepsilon _L  = \left( {\begin{array}{*{20}c}
   {\varepsilon _ \bot  \sin ^2 \alpha  + \varepsilon _\parallel  \cos ^2 \alpha } & {\Delta \varepsilon \sin \alpha \cos \alpha } & 0  \\
   {\Delta \varepsilon \sin \alpha \cos \alpha } & {\varepsilon _ \bot  \cos ^2 \alpha  + \varepsilon _\parallel  \sin ^2 \alpha } & 0  \\
   0 & 0 & {\varepsilon _ \bot  }  \\
\end{array}} \right)
\end{equation}
%
%
\begin{equation}
\varepsilon _R  = \left( {\begin{array}{*{20}c}
   {\varepsilon _ \bot  \sin ^2 \beta  + \varepsilon _\parallel  \cos ^2 \beta } & 0 & {\Delta \varepsilon \sin \beta \cos \beta }  \\
   0 & {\varepsilon _ \bot  } & 0  \\
   {\Delta \varepsilon \sin \beta \cos \beta } & 0 & {\varepsilon _ \bot  \cos ^2 \beta  + \varepsilon _\parallel  \sin ^2 \beta }  \\
\end{array}} \right)
\end{equation}
where $\Delta \varepsilon  = \left( {\varepsilon _\parallel   - \varepsilon _ \bot  } \right)$ is the birefringence and $\alpha $  and $\beta $  are the angles that the two optical axes form with the $x$ axis on the $(x,z)$ and $(x,y)$ planes respectively (see Fig. 1).
The rotation of $\alpha $   the optical $C$-axis on the $(x,z)$ makes asymmetric the two like-Dirac points due to the $z$-component of the electric field (see Fig. 5(a)), while the rotation of  $\beta $  on the plane $(x,y)$ opens an independent from polarization photonic gap  between the two polarized bands (see Fig. 5(b)) and, moreover, by tuning its value (in the present case $\alpha  = 45^\circ $) a SIP appears in the dispersion curve close to the $\hbar \omega  = 1.5887eV$  photon energy (see Fig. 5(b)); therefore under the former conditions also the hybrid photonic crystals become optically active.  
Notice that the tailoring of the $N$-cluster requires a value of  $N$ as large as to show  well formed optical gaps (see Fig. 5(b)), but not so large to give any small details,  present in the dispersion curves, that are not interesting in the present study of spin-orbit effect on the exciton-polariton propagation in resonant hybrid photonic clusters.   For the present used parameter values a sensible choice is for $N = 34$ as already discussed in Ref. \cite{3} where also an  easy   method,  based on reflectivity optical response, for obtain energy resonance between exciton transition in a 2D-quantum wells  and the energy of the SIP, of the corresponding resonant hybrid photonic crystal, is  discussed.  
Let us consider a symmetric envelope function $(\Psi )$ of Wannier light-hole exciton, performed by a product of two sub-bands, with transition energy in resonance with the former SIP point of dispersion curves  $(E_{ex} (\vec K_\parallel   = 0.0) = 1.588eV)$.

The change of the dispersion curves of Fig. 5(a) and 5(b), due to the former exciton-polariton interaction, and close to the exciton energy are shown in Figs. 6(a) and 6(b) respectively. Notice that the exciton-polariton dispersion curves under resonant condition between exciton transition energy and the energy of the SIP point, of the photonic crystal, are strongly asymmetric with respect to forward-backward propagation (see Fig. 6(b)). Moreover, while the asymmetric shape of the dispersion curves shown in Figure 6a does retains the rotational symmetry observed in Figure 4,  differently  this later is strongly modified by the presence of a SIP point in the dispersion curves. 
In fact let us consider the optical response of non-normal and TE polarized  incidence wave in an $N$-cluster of elementary cells for parameter values of  Fig. 6(b). Notice that, in this case, the TE energy gap results at photon energy  higher than the,  independent from the polarization, optical gap. 
In Figs. 7(a) and 7(b) the ratios $I_z /I_\parallel  $   are computed under resonance with the SIP exciton energy $(E_{ex}  = 1.588eV)$ as a function of rotation angle $\varphi $  and for  two different values of exciton-spin-orbit interaction, namely: $\beta _{ex}  = 0.7eV$\AA  and $\beta _{ex}  = 0.14eV$\AA (red squares) respectively. In the same figures are also reported for comparison  the $I_z /I_\parallel  $  values obtained in absence of SIP (blue circles).  Moreover, in Fig. 7(a)  the results obtained for an exciton transition energy $E_{ex}  = 1.6eV$   out-of-resonance  with respect to the SIP  energy (green triangles) is also shown. 
Notice that for both the $\beta _{ex} $ values and in  both the cases (in and out of resonance) the curves appear deformed  with respect to the symmetrical one; furthermore, they   do not  become negligible for odd multiples of $\pi /4$    and the initial $\pi /2$ periodicity  of the spin-orbit Hamiltonian with respect to the in-plane rotation $\varphi $  of the semiconductor elementary cell,  appears doubled.
Moreover,  for symmetrical forward-backward polariton propagation $\beta _{ex}  = 0.14eV$\AA  an almost circular pattern (red squares in Fig. 7(b)) highlights  an almost independence of spin-orbit effect from the orientation of the semiconductor elementary cell.

Notice that the asymmetry observed in the exciton-polariton curves of Fig.4a of ref.(2), with respect to the symmetrical one of Fig.4b ,  gives in Fig.4 of the
present work  a change, from red square curve to blue circle curve, that underlines the contribution due to the optical activity (the spin-orbit effect) of the 
light hole exciton polariton .

\section{CONCLUSION}
In conclusion, for excitons both in  and out-of-resonance with the energy of the SIP,  it is always obtained  a strong deformation of the polar curves of the optical activity,  as already described in Ref. \cite{3} where an  interaction parameter value $\beta _{ex}  = 0.7eV$\AA was used. Finally,  for the here used value $\beta _{ex}  = 0.14eV$\AA the polar curve of the optical activity appears almost independent of the rotation angle $\varphi $   (see Fig. 7(b)) and this unexpected behavior is the most interesting result of the present work.
\section*{Acknowledgments}
The Author is indebted with Dr. L. Pilozzi and Dr. N. Tomassini for helpful discussions.
%
%

%
\end{document}